\documentclass[fleqn,usenatbib]{mnras}
\usepackage[T1]{fontenc}

\DeclareRobustCommand{\VAN}[3]{#2}
\let\VANthebibliography\thebibliography
\def\thebibliography{\DeclareRobustCommand{\VAN}[3]{##3}\VANthebibliography}

\usepackage{graphicx}	
\usepackage{amsmath}	
\usepackage{amssymb}	
\usepackage{soul}
\usepackage{xcolor}
\DeclareRobustCommand{\insig}[1]{{\textcolor{gray}{#1}}}
\usepackage[multiple]{footmisc}
\usepackage{newtxtext,newtxmath}

\newcommand{\mcs}[1]{\multicolumn{1}{c}{#1}}
\newcommand{\mcd}[1]{\multicolumn{2}{c}{#1}}

\title[Meteor shower radiant dispersions]{Meteor shower radiant dispersions in Global Meteor Network data}

\author[A. V. Moorhead et al.]{
Althea V. Moorhead,$^{1}$\thanks{E-mail: althea.moorhead@nasa.gov}
Tiffany Clements,$^{2}$
and Denis Vida$^{3}$
\\
$^{1}$NASA Meteoroid Environment Office, Marshall Space Flight Center EV44, Huntsville, Alabama 35812, USA\\
$^{2}$Aerodyne Industries, Jacobs Space Exploration Group, Marshall Space Flight Center EV44, Huntsville, Alabama 35812, USA\\
$^{3}$Department of Physics and Astronomy, University of Western Ontario, London, Ontario, N6A 3K7, Canada
}

\date{Accepted XXX. Received YYY; in original form ZZZ}

\pubyear{2021}

\begin{document}
\label{firstpage}
\pagerange{\pageref{firstpage}--\pageref{lastpage}}
\maketitle

\begin{abstract}
Meteor showers occur when streams of meteoroids originating from a common source intersect the Earth. 
There will be small dissimilarities between the direction of motion of different meteoroids within a stream, and these small differences will act to broaden the radiant, or apparent point of origin, of the shower. This dispersion in meteor radiant can be particularly important when considering the effect of the Earth's gravity on the stream, as it limits the degree of enhancement of the stream's flux due to gravitational focusing. In this paper, we present measurements of the radiant dispersion of twelve showers using observations from the Global Meteor Network. We find that the median offset of individual meteors from the shower radiant ranges from 0.32$^\circ$ for the eta Aquariids to 1.41$^\circ$ for the Southern Taurids. We also find that there is a small but statistically significant drift in Sun-centered ecliptic radiant and/or geocentric speed over time for most showers. Finally, we compare radiant dispersion with shower duration and find that, in contrast with previous results, the two quantities are not correlated in our data.
\end{abstract}



\section{Introduction}

Meteor showers are often conceptualized as streams of meteoroids that encounter the Earth on parallel paths. Meteoroids with identical velocity vectors can be described by a single radiant (apparent point of origin) and speed. In actuality, the particles in a meteoroid stream will have some non-zero dispersion in their radiants and speeds for a variety of reasons: meteoroids are ejected from their parent comets at varying times and with varying speeds, directions, and material properties, and these differences are amplified over time as the meteoroids are exposed to solar radiation and planetary perturbations \citep{2019msme.book..161V}.

\cite{2020MNRAS.494.2982M} demonstrated that characterizing radiant dispersion is vital for a realistic computation of the effects of a planet's gravity on a meteoroid stream. When zero dispersion is assumed, a singularity in meteoroid flux and number density appears along the anti-radiant line (that is, locations in space opposite the planet from the radiant). When the radiant dispersion is incorporated into the gravitational focusing computation, this singularity disappears and the flux and number density are finite, with the magnitude of the flux along the anti-radiant line depending on the radiant dispersion. Thus, radiant dispersion measurements are necessary to correctly predict the meteoroid flux onto high-orbiting spacecraft or even the Moon \citep{2020MNRAS.494.2982M}. It is particularly important to separate measurement error from the true radiant dispersion, as overestimates in radiant dispersion produce underestimates in flux.

Meteor shower radiant dispersions probe the degree of orbital dissimilarity within a meteoroid stream. Another measure of orbital dissimilarity -- the width and shape of a shower's activity profile -- is often used to calibrate or test stream models \citep{2019Icar..330..123E}. Thus, the radiant dispersion could be used as another constraint on the age and/or dynamical history of a meteoroid stream \citep{1998Icar..133...36B}, particularly if it reflects a non-symmetrical stream cross-section, as we discuss in Section~\ref{sec:dur}.

Radiant dispersions are not, however, routinely measured. In many cases this is because the precision of measured meteor trajectories does not permit a useful characterization \citep{2019msme.book...90K}. For instance, early studies attempted to measure radiant dispersions using single-station meteor observations \citep{1950ApJ...111..104J,1954MNRAS.114..229W,1963mmc..book..674M}; their methodology involved fitting great circles to the meteors' paths, determining the point of closest convergence, and then measuring the minimum displacement from this point to each meteor arc. This approach places a lower limit on the individual radiant offsets and thus likely underestimates the radiant dispersion.

Multi-station meteor observations permit the calculation of radiant points rather than arcs and, at least in theory, enable better measurements of radiant dispersion. \cite{1970BAICz..21..153K} used double-station photographic observations of meteors to measure the average or median angular deviations of meteor radiants from the shower average. The authors combined data from multiple sources: the Harvard \citep[most notably the Super-Schmidt cameras;][]{hawkins1961orbital, mccrosky196825}, Ond\v{r}ejov \citep{ceplecha1964ondrejov}, and Soviet meteor observation programs \citep{babadzhanov1967orbits}. As a result, the data were not reduced using a consistent method. Furthermore, the data set did not contain uncertainties on individual radiant measurements and thus no quality thresholds could be imposed. Although their Perseid radiant dispersion is based on a large number of meteors, they had to rely on a relatively small number of meteors for most showers (less than 50 in all but two cases: the Perseids and the Geminids). No other studies systematically survey meteor shower radiant dispersions, but the literature contains scattered characterizations of the dispersion of individual showers \citep[e.g.,][]{1997A&A...317..953J,1998MNRAS.301..941J}.

This paper presents the first survey of meteor shower radiant dispersions conducted in the past 50 years. We characterize the Sun-centered geocentric ecliptic radiant drift and radiant dispersion of twelve meteor showers using high-quality video meteor trajectories collected by the Global Meteor Network \citep{vida2021global}. We use a bootstrap approach to compute uncertainties in our fit parameters and attempt to account for the systematic increase in radiant dispersion caused by radiant measurement errors. We find that radiant dispersion tends to shrink as we restrict our solutions to shorter periods of activity. Finally, we compare our measured dispersions with the shower durations in an attempt to reproduce the correlation observed by \cite{1954MNRAS.114..229W}.

\section{Data}
\label{sec:data}

\subsection{Global Meteor Network}
\label{sec:gmn}

The meteor trajectory data used in this work were collected by the Global Meteor Network (GMN)\footnote{\url{https://globalmeteornetwork.org/}} between December 2018 and May 2021; about 300 of the more than 450 GMN cameras contributed \citep[see][for a map of contributing cameras]{vida2021global}. GMN video cameras are operated by professional organizations as well as amateur astronomers and citizen scientists; all trajectory data are made publicly available in near-real time. 

Because GMN relies on volunteers to build and operate its camera stations, the specifics of individual cameras can vary. In this work, we use data from cameras with two different sets of specifications: cameras with a 3.6-mm focal length, $88^\circ \times 48^\circ$ field of view, and a limiting stellar magnitude of about +6; and high-resolution cameras with a 16-mm focal length, $20^\circ \times 10^\circ$ field of view, and a limiting stellar magnitude of about +9 \citep[additional details are provided in Table 1 of][]{2020MNRAS.494.2982M}. A detailed description of the network, the systems and the methodology used by GMN is given in \cite{vida2021global}.

\subsection{Trajectory solutions}

The trajectory solutions for each meteor were calculated using the observational Monte Carlo (MC) approach detailed in \citet{vida20a} and \citet{vida2021global}. This technique uses the result provided by the intersecting planes method \citep{ceplecha87} as an initial solution for trajectory minimization using the line-of-sight method developed by \citet{borovicka90}. This method calculates 3D meteor trajectories by simultaneously using optical observations from multiple stations. The measurement error is characterized as angular residuals between the measurements and the fitted trajectory. Gaussian-distributed noise on the order of the determined errors is added to the original measurements and the trajectory is recomputed in every MC run. This gives a purely geometrical trajectory fit uncertainty. The novelty of the Monte Carlo method is that it utilizes the fact that all observers must observe the same meteor dynamics (velocity and deceleration) at the same time, thus a cost function based on residuals of time vs.\ distance from the meteor beginning is used to choose the best solution. This results in a final trajectory solution which is optimal both geometrically and dynamically, and provides realistic uncertainties for every orbital parameter. A similar approach was taken by \citet{gural2012solver}, but their method was forcing an unphysical model of meteor kinematics to observations, leading to biases in the computed trajectory solution \citep{egal2017challenge}. Nevertheless, we note that without full meteor ablation and fragmentation modeling, measured initial velocities of meteors can be underestimated by as much as 750~m~s$^{-1}$ \citep{vida2018modelling}.

\subsection{Quality cuts}

Our 30-month span of GMN meteor data contains more than 200,000 trajectories, about 40\% of which are tentatively associated with a meteor shower. 
We restrict our analysis to the most precise trajectories by applying similar quality cuts as were used in \cite{2020MNRAS.494.2982M}. First, we require that the maximum convergence angle between camera lines of sight is at least 15$^\circ$. Second, we require that the median angular fit error is no larger than 1~arcmin. Finally, we require that the formal uncertainty in the average in-atmosphere speed is no greater than 1~km~s$^{-1}$, and that the formal uncertainty in the radiant is no greater than 0.25$^\circ$. These cutoffs allow us to exclude meteors with the largest errors while retaining a substantial proportion of the data.

Ideally, we would also restrict our analysis to meteors observed by more than two stations and/or by high-resolution cameras with 16-mm lenses. However, this would significantly reduce the number of meteors available to us. We therefore refrain from completing an analysis using high-resolution or many-station radiant solutions, but note that \cite{2020MNRAS.494.2982M} found that the radiant offset distribution of the high-resolution cameras was similar to that measured using all cameras.

\section{Methods}
\label{sec:methods}

\subsection{Shower member selection and radiant drift}
\label{sec:drift}

Meteor shower radiants are known to drift in geocentric right ascension and declination over time. However, this drift is minimized when the radiant is rotated into ecliptic coordinates and the Sun's ecliptic longitude subtracted from the shower longitude \citep{2008Icar..195..317B,2017P&SS..143..116J}. The resulting Sun-centered geocentric ecliptic coordinates are usually denoted $\lambda_g - \lambda_\odot$ and $\beta_g$. We will use Sun-centered ecliptic coordinates to describe the meteor radiants considered in this paper. The subscript $g$ indicates that these radiants are geocentric (rather than heliocentric). However, while the Sun-centered ecliptic coordinate system is known to minimize radiant drift, we do not assume that it eliminates all drift. Instead, we also test and fit for any drift in Sun-centered ecliptic radiant and speed.

First, we select meteors that lie near a shower's established solar longitude, radiant, and speed. Typically, one would complete this step by selecting meteors whose radiants lie within a certain great-circle distance of the shower radiant, or whose orbital elements satisfy some orbital similarity criterion \citep[e.g.,][]{1981Icar...45..545D}. However, in order to minimize any biases in fitting for linear drift in the individual radiant angles, we instead select meteors falling within the deliberately large, quasi-rectangular region where 
\begin{align}
| \beta_g - \beta_\mathrm{ref} | &< 10^\circ \, , 
    \label{eq:db0} \\
|\Delta \lambda_g - \Delta \lambda_\mathrm{ref}| \cdot \cos \beta_\mathrm{ref} &< 10^\circ \, , 
    \label{eq:dl0} \\
| v_g - v_\mathrm{ref} |/v_\mathrm{ref} &< 0.2 \, \textrm{, and} 
    \label{eq:dv0} \\
| \lambda_\odot - \lambda_0 | &< \tfrac{1}{2}\mathrm{dur}_\mathrm{max} .
    \label{eq:ds0}
\end{align}
Here, we use $\Delta \lambda_g$ as shorthand for $\lambda_g - \lambda_\odot$. We initially centered these windows about the reference solar longitude, radiant, and speed reported by \cite{2009JIMO...37...55S}, but subsequently shifted them to center around the median solar longitude of the shower members and the corresponding best-fit radiant and speed at that point in time (see Table~\ref{tab:drift}).

The variable $\mathrm{dur}_\mathrm{max}$ is the estimated duration of the shower. Ideally, one would choose a very short duration in order to probe the radiation dispersion of the ``core'' of the stream. However, for long-lasting, low-activity showers such as the Taurids, a short duration selects far too few shower members. For each shower, we ran our shower membership algorithm for values of $\mathrm{dur}_\mathrm{max}$ ranging from 1~day ($\sim0.986^\circ$) to 29~days ($\sim28.583^\circ$) in increments of 2~days. We then selected as our nominal value of $\mathrm{dur}_\mathrm{max}$ the duration that yielded at least 60\% of the estimated true shower members (see Section~\ref{sec:spor} for our sporadic contamination estimate) yielded by the 29~day case. This ranged from 1~day for the Quadrantids to 19~days for the Southern Taurids.

Within our selection window, we use least-squares minimization to derive linear fits in our parameters. We express these drifts as:
\begin{align}
    \Delta \lambda_\mathrm{ref} (\lambda_\odot) &= m_\lambda \cdot (\lambda_\odot - \lambda_0) + b_\lambda \label{eq:driftl} \\
    \beta_\mathrm{ref} (\lambda_\odot) &=
    m_\beta \cdot (\lambda_\odot - \lambda_0) + b_\beta \label{eq:driftb} \\
    v_\mathrm{ref} &= m_v \cdot (\lambda_\odot - \lambda_0) + b_v \label{eq:driftv}
\end{align}
As discussed in appendix~\ref{sec:resid1}, we clip our data to exclude the outliers caused by sporadic contamination. We accomplish this by performing an initial fit using all data, then removing points that lie more than 2.5$^\circ$ from this best-fit radiant (see equation~\ref{eq:bigtheta} below), and then perform a second least-squares fit using the remaining meteors. We bootstrap this process 1000 times and derive a covariance matrix (and errors) from the bootstrap results.

\begin{figure}
    \raggedright
    \includegraphics[]{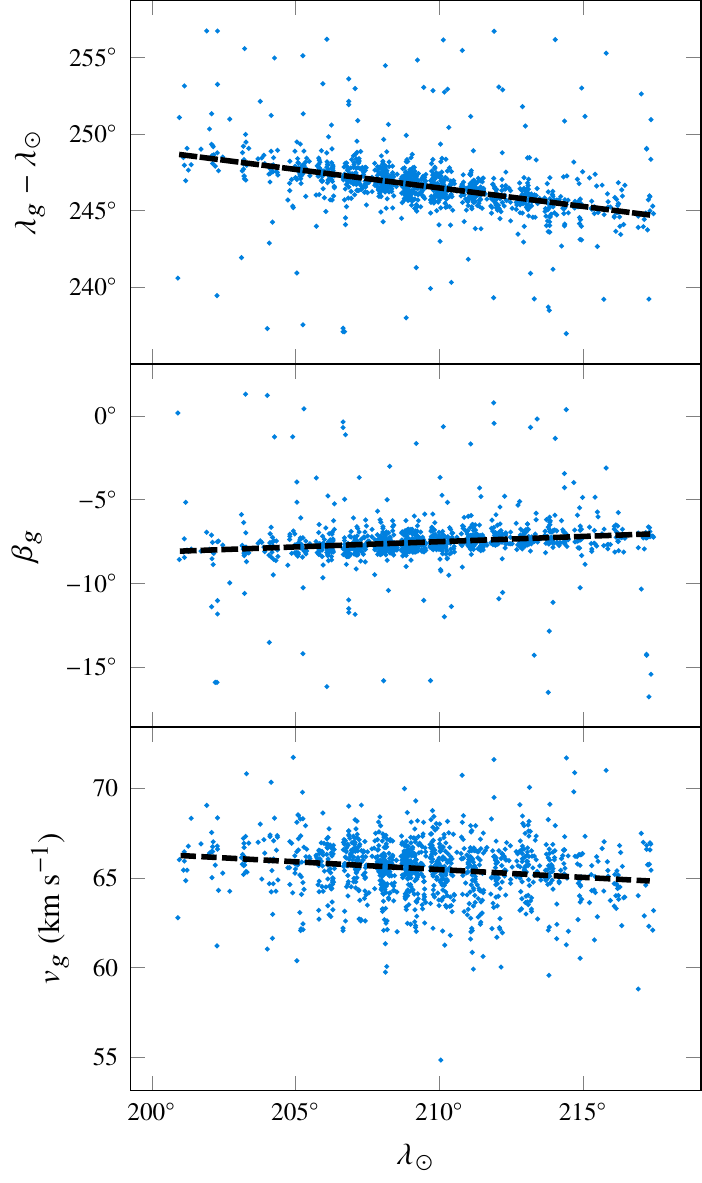}
    \caption{Sun-centered ecliptic longitude (top), latitude (center), and geocentric speed (bottom) plotted against solar longitude for meteors observed by GMN in the vicinity of the Orionid meteor shower. A linear fit to the data is shown as a dashed black line. Plots of the residuals can be found in Fig.~\ref{fig:linear_resids}.}
    \label{fig:linear_fits}
\end{figure}

Figure~\ref{fig:linear_fits} shows sample linear drift fits for the Orionid meteor shower. In this case, we find that there is a statistically significant correlation between solar longitude and $\lambda_g - \lambda_\odot$, $\beta_g$, and $v_g$, but the drift in $\lambda_g - \lambda_\odot$ is the most pronounced, with a value of $-0.22^\circ$ per degree in solar longitude. It is also apparent from Fig.~\ref{fig:linear_fits} that our chosen angular window of $\pm 10^\circ$ is much larger than the typical spread in radiant angles for this shower.

Once we have measured the drift in shower radiant and speed, we re-select shower member candidates as follows:
\begin{align}
    \Theta \left( \Delta \lambda_g, \, \beta_g, \, \Delta \lambda_\mathrm{ref}(\lambda_\odot), \, \beta_\mathrm{ref} (\lambda_\odot) \right) &< 10^\circ \label{eq:bigtheta} \\
    | v_g - v_\mathrm{ref} (\lambda_\odot) |/v_\mathrm{ref} &< 0.2 \, \textrm{, and} \label{eq:dv1} \\
| \lambda_\odot - \lambda_0 | &< \tfrac{1}{2} \mathrm{dur}_\mathrm{max} \, . \label{eq:ds1}
\end{align}
where $\Theta$ is the central angle\footnote{computed using the haversine formula} between the two longitude-latitude pairs ($\Delta \lambda_g$, $\beta_g$) and ($\Delta \lambda_\mathrm{ref} (\lambda_\odot)$, $\beta_\mathrm{ref} (\lambda_\odot)$) and our reference radiant and speed now varies with solar longitude according to equations~\ref{eq:driftl}-\ref{eq:driftv}.

\subsubsection{Special handling of the Taurid complex}
\label{sec:tau}

We wished to include the Northern and Southern Taurid (NTA and STA) meteor showers in our analysis, but this pair of unusual showers requires special handling. The Taurids have relatively low peak activity but are very long-lasting compared to other showers; activity persists for several months rather than days or weeks. They are embedded in the center of the anti-helion sporadic source and may arise from the same parent body, 2P/Encke \citep{Whipple40,2009Icar..201..295W}. While they are considered meteor showers, the Taurids appear to lie somewhere between meteor shower and sporadic source in their characteristics. Furthermore, because the Nothern and Southern Taurids are two branches of the same complex, have similar orbits, and are active at the same time and with similar radiants, most shower identification techniques struggle to separate the two branches \citep[see, for example,][]{2017M&PS...52.1048S}. 

\begin{figure}
    \includegraphics{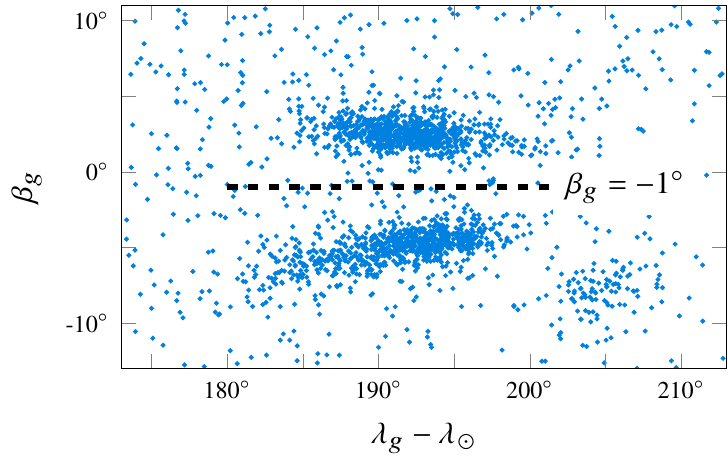}
    \caption{Sun-centered ecliptic meteor radiants in the vicinity of the Northern and Southern Taurid meteor showers. We plot only those meteors that occur between $\lambda_\odot = 200$ and $\lambda_\odot = 250$. The $\beta_g = -1^\circ$ dividing line is marked with a dashed line.}
    \label{fig:taurid}
\end{figure}

We have found that the simplest and most straightforward way to separate the two showers is by ecliptic latitude (see Fig.~\ref{fig:taurid}). We consider only those meteors with $\beta_g > -1^\circ$ when characterizing the NTA radiant distribution, and only those meteors with $\beta_g < -1^\circ$ when characterizing the STAs. This choice is also applied to the sporadic contamination estimate (see Section~\ref{sec:spor}). 

\subsection{Sporadic contamination}
\label{sec:spor}

It is impossible to tell which of the meteors in our radiant-speed-solar longitude window are truly shower members and which are false positives; no shower membership selection method can avoid contamination by sporadic meteors. We therefore estimate the distribution of sporadic contaminants by selecting meteors that lie near our reference shower radiant and speed at times when the shower is \emph{not} active. We accomplish this by shifting the solar longitudes of all meteors by 60$^\circ$, 63$^\circ$, and so on up to 300$^\circ$ from their measured values and rerunning our shower membership algorithm. This allows us to assess the distribution of sporadic meteoroids around the drifting shower radiant, averaging over an eight-month period when the shower is not active in order to minimize the impact of effects such as gaps in observations.

\begin{figure}
    \centering
    \includegraphics[]{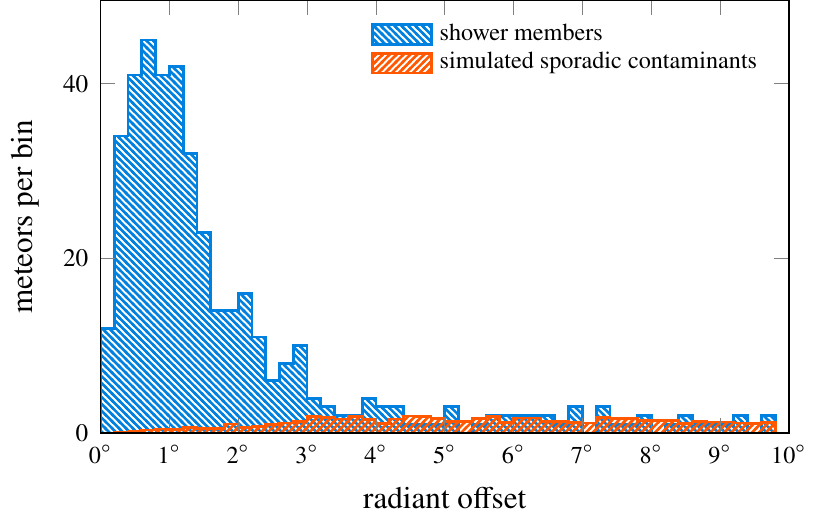}
    \caption{The distribution of radiant offsets (relative to the shower reference) of GMN meteors that satisfy equations~\ref{eq:bigtheta}-\ref{eq:ds1} for the Northern Taurid (NTA)  meteor shower (``shower members''), and those that satisfy these equations after being shifted in solar longitude (``simulated sporadic contaminants'').}
    \label{fig:rawhist}
\end{figure}

Figure~\ref{fig:rawhist} displays the angular distribution of these simulated sporadic meteors about the drifting shower radiant for the Northern Taurid meteor shower. We compare these simulated sporadic contaminants with the angular distribution of unmodified meteors about the same shower radiant. Note that while the number of shower members per bin is an integer number, the number of sporadic contaminants is not necessarily an integer; this is because we have divided the latter by the number of times we have shifted the data in solar longitude. Based on these results, we expect that approximately 15\% of the nominal shower members are actually sporadic contaminants (we later fit for the level of sporadic contamination and obtain a value of 12\%; see Section~\ref{sec:fit} and Table~\ref{tab:nomfits}). This fraction ranges from a fraction of a percent for the Geminids (GEM) to nearly 20\% for the alpha Capricornids (CAP). 

We consider this sporadic contamination estimation when selecting which showers to characterize. In this paper, we analyze showers for which the estimated number of showers members within our nominal solar longitude window is greater than 100 after subtracting the estimated sporadic contamination level.

However, it is important to note that these are only rough estimates for the contamination level. Sporadic activity can vary seasonally \citep{2006MNRAS.367..709C,2009M&PS...44.1837C} and therefore we allow the sporadic contamination level to vary in our fits; see Section~\ref{sec:fit}. The true value of generating simulated sporadic meteors in the manner described here is that it allows us to characterize the shape of the sporadic radiant distribution near ecliptic radiants that are similar to a given meteor shower.

\subsection{Distribution fitting}
\label{sec:fit}

Because we expect our data to be contaminated by sporadic meteors, we must incorporate that assumption into our fitting algorithm. We do this by constructing histograms of the angular offsets between our drifting shower radiant and both the nominal shower members and our simulated sporadic data. We then assume that the distribution of true shower members, $f_i$, is a linear combination of the nominal shower members ($h_i$) and the simulated sporadic data ($g_i$):
\begin{align}
    f_i &= (1+c) h_i - c g_i \label{eq:fi}
\end{align}
Here, $h_i$ and $g_i$ are assumed to be the fraction of meteors per bin or per degree; the choice of coefficients ensures that this is also true for $f_i$. Section~\ref{sec:sigma} discusses the uncertainties associated with these quantities in detail. The unitless parameter $c$ is a measure of the degree of sporadic contamination. Because we include $c$ in our fitting process, we do not require the sporadic activity level (or observational coverage) during the shower to be equal to that when the shower is not active; in fact, the activity of the sporadic sources is known to vary seasonally \cite{2006MNRAS.367..709C,2009M&PS...44.1837C}. However, our approach does assume that the distribution of sporadic radiants about the shower's sun-centered ecliptic geocentric radiant is constant.

We first fit a Rayleigh distribution to our data; if the vectorial components of the meteoroids' velocity vectors are Gaussian, the radiant dispersion can be approximated as a Rayleigh distribution \cite[see Appendix A of][]{2020MNRAS.494.2982M}. We assume that the minimum offset is $0^\circ$ and that the only free parameter is the mode of the distribution, $\mu$. The probability density function is given by:
\begin{align}
    \mathrm{PDF}_1(\mu, \Theta) &= \frac{\Theta}{\mu^2} e^{-\Theta^2/(2 \mu^2)}
    \label{eq:pdf1}
\end{align}
However, our histogram bins have finite width, and we wish to take this into account in performing our fits. Therefore, we fit the following function to our data:
\begin{align}
    \mathrm{CDF}_1(\mu, \Theta_{i+1}) - \mathrm{CDF}_1(\mu, \Theta_i)
\end{align}
where $\Theta_i$ and $\Theta_{i+1}$ mark the edges of a given bin, and CDF is the cumulative distribution function:
\begin{align}
    \mathrm{CDF}_1(\mu, \Theta) &= 1 - e^{-\Theta^2/(2 \mu^2)}
    \label{eq:cdf1}
\end{align}
Our nominal bin width is 0.2$^\circ$, but we later include bin width as a potential source of uncertainty (see Section~\ref{sec:bootstrap}).

\begin{figure*}
    \centering
    \includegraphics{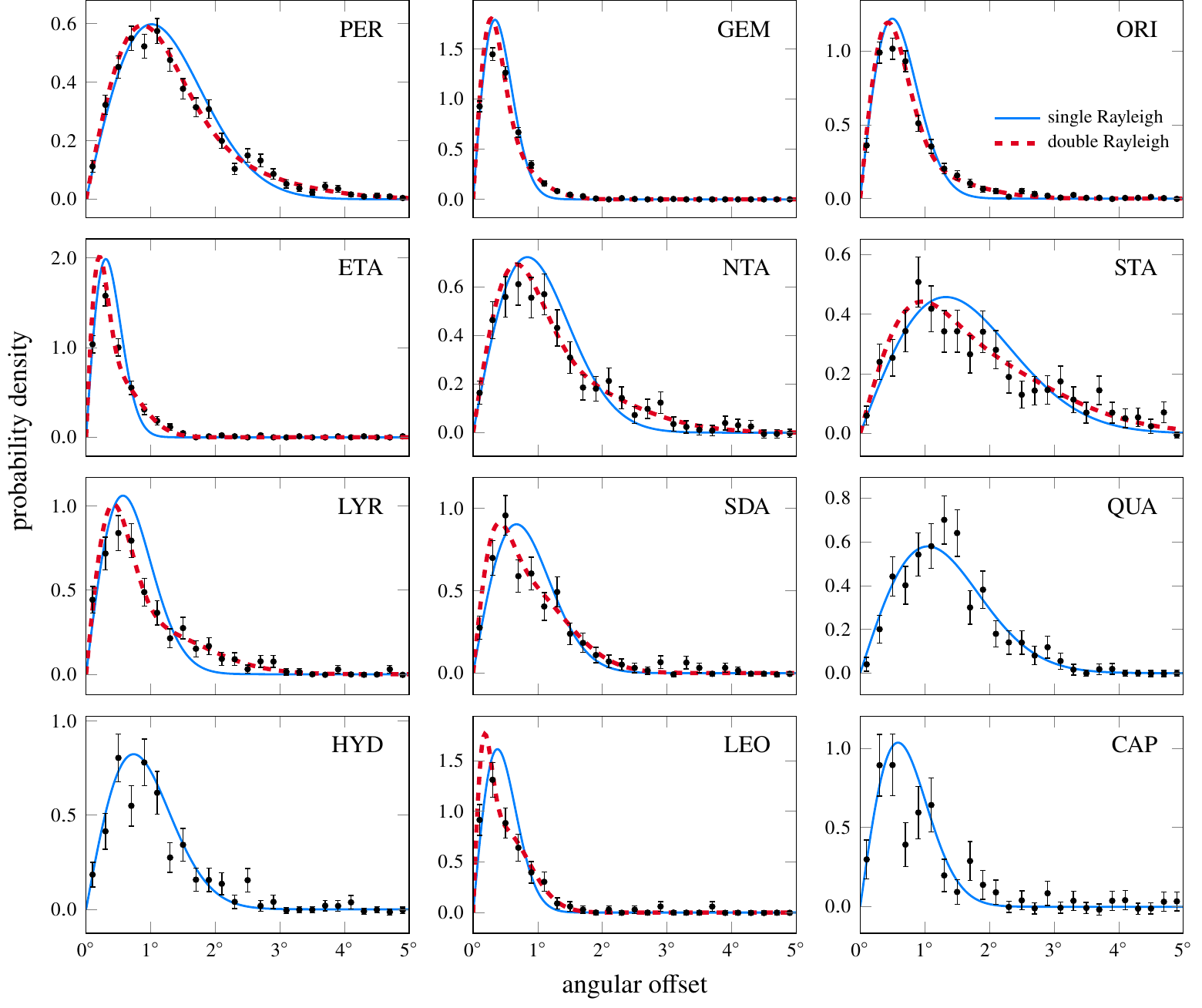}
    \caption{The noise-subtracted distribution of radiant offsets measured by GMN for twelve meteor showers (black points). For each shower, we also display the best-fitting Rayleigh distribution (solid blue line) and, if applicable, double Rayleigh distribution (dashed red line).}
    \label{fig:pdfs}
\end{figure*}

In many cases, we found that we could not fully replicate the shape of the data using a Rayleigh distribution. In these cases, we fit a double Rayleigh distribution to the data, whose probability and cumulative distribution functions are:
\begin{align}
    \mathrm{PDF}_2 &= (1 - \gamma) \, \mathrm{PDF}_1(\mu_a, \Theta) +
        \gamma \, \mathrm{PDF}_1(\mu_b, \Theta) \label{eq:pdf2} \\
    \mathrm{CDF}_2 &= (1 - \gamma) \, \mathrm{CDF}_1(\mu_a, \Theta) +
        \gamma \, \mathrm{CDF}_1(\mu_b, \Theta) \label{eq:cdf2}
\end{align}
where $\gamma$ is a unitless parameter that measures the contribution of the second Rayleigh distribution to the overall distribution.
Figure~\ref{fig:pdfs} compares both distributions -- single and double Rayleigh -- with the noise-subtracted distribution of radiant offsets measured for nine showers. In most cases, the data have a shallower slope on the right side of the peak than a single Rayleigh distribution predicts, and thus are better fit by a double Rayleigh distribution (see Section~\ref{sec:resid2} for an analysis of the fit residuals). The Quadrantids (QUA), sigma Hydrids (HYD), and alpha Capricornids (CAP) are exceptions; use of a double Rayleigh results in no substantial decrease in the reduced chi-squared statistic (see table~\ref{tab:chi2red}). However, these three showers have some of the poorest number statistics out of the set we analyze. Therefore, throughout this paper we order showers by the number of meteors included in the nominal fit (see Table~\ref{tab:nomfits}), so that more reliable fits are located at the top of figures and tables.

We find the best fit Rayleigh or double Rayleigh distribution by minimizing the chi-squared statistic, an approach that will be very familiar to most readers. However, we would like to note that one of our fit parameters, $c$, is not contained within the Rayleigh and double Rayleigh distributions discussed above. Therefore, we minimize the following quantity:
\begin{align}
    \chi^2 &= \sum_i{\frac{
        \left( \mathrm{CDF}(\Theta_{i+1}) - \mathrm{CDF}(\Theta_i) 
        - [(1+c) h_i - c g_i] \right)^2}{
        (1+c)^2 \, \sigma_{h_i}^2 + c^2 \, \sigma_{g_i}^2}}
\end{align}
In the case of a single Rayleigh distribution, we fit two parameters: $\mu$ and $c$. In the case of a double Rayleigh distribution, we fit four parameters: $\mu_a$, $\mu_b$, $\gamma$, and $c$.

\subsection{Uncertainties and sensitivity analysis}
\label{sec:uncert}

\subsubsection{Uncertainty estimation}
\label{sec:sigma}

We have chosen to fit Rayleigh distributions to a histogram of our data values. Histogram values are sometimes assigned uncertainties of $\sigma \approx \sqrt{n_i}$, where $n_i$ is the number of data points that fall within bin $i$; this is the root of the Poisson variance $n_i$. However, the probability that a value falls into a given bin is better described by a binomial distribution. If the probability that a meteor falls within bin $i$ is $p_i$ and the total number of observed meteors is $N$, the variance is $p_i (1 - p_i) / N$ under the normal approximation to a binomial distribution. When $p_i N = n_i$ is small, this approximation in turn breaks down. 

We therefore use half the width of a one-$\sigma$ Wilson-Agresti-Coull \citep[WAC;][]{Agresti1998,Ott2016} confidence interval to compute the uncertainty associated with our histogram bins:
\begin{align}
    \sigma_{h_i} &\simeq \sqrt{\tilde{h}_i (1 - \tilde{h}_i) / (N + 1)}, \, \text{where} \label{eq:wac} \\
    \tilde{h}_i &= \left( n_i + \tfrac{1}{2} \right)/(N + 1)
\end{align}
This approximation allows us to include $n_i = h_i = 0$ bin values in our chi-squared minimization. (We do not, however, replace $h_i$ with $\tilde{h}_i$, as this would artificially inflate the tail of the distribution.)

As discussed in Section~\ref{sec:fit}, we fit Rayleigh distributions to noise-subtracted data, $f_i$ (see equation~\ref{eq:fi}). The uncertainty associated with these data is therefore:
\begin{align}
    \sigma_{f_i} &= \sqrt{(1+c)^2 \, \sigma_{h_i}^2 + c^2 \, \sigma_{g_i}^2} \, .
    \label{eq:wtot}
\end{align}

\subsubsection{Bootstrap uncertainties}
\label{sec:bootstrap}

\begin{figure*}
    \centering
    \includegraphics{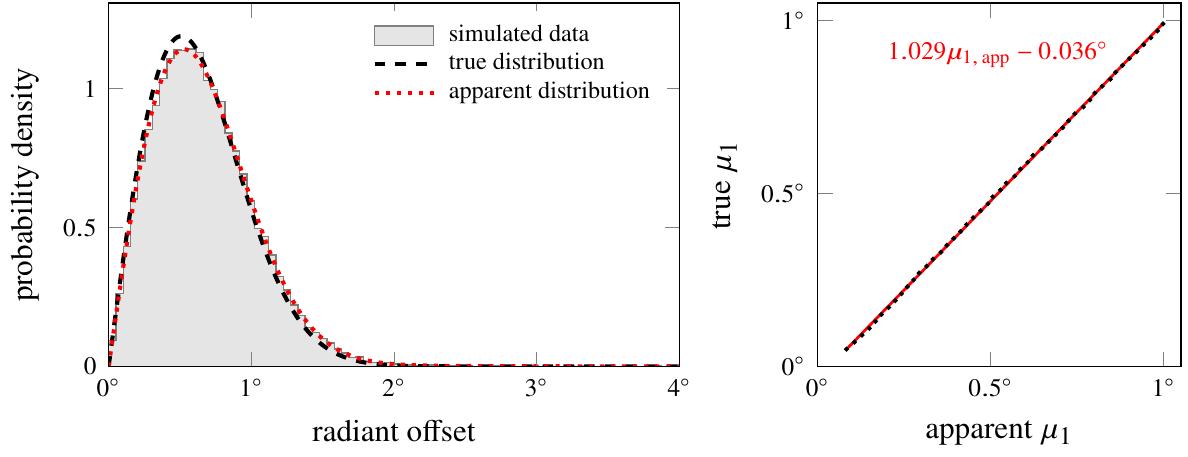}
    \caption{Distribution of simulated meteor radiants (left) following a Rayleigh distribution with a mode of 0.51$^\circ$ and uncertainties in $\lambda_g$ and $\beta_g$ taken from the observations. The right panel compares the true underlying spread in radiant (true $\mu_1$) with the apparent spread in radiant (apparent $\mu_1$).}
    \label{fig:spread}
\end{figure*}

Many fitting routines either return parameter uncertainties or a covariance or inverse Hessian matrix. However, because our algorithm contains multiple steps (a least-squares drift fit followed by a radiant offset distribution fit), we estimated our uncertainties using a classic bootstrap approach. From our initial sample of meteors satisfying equations~\ref{eq:db0}--\ref{eq:ds0}, we generated 1000 random samples of the same size, sampled with replacement. Simultaneously, we randomly vary the radiant and geocentric velocity drift using the covariance matrix corresponding to the best fit. Each bootstrap sample was subjected to our fitting algorithm. We then computed the difference between the 16th and 84th percentiles of each bootstrapped parameter (drift slopes, Rayleigh modes, noise level, etc.) and assume that this is equivalent to twice the standard deviation in each case.

The above process allows us to characterize the uncertainty in our fitting process. However, we also make a number of somewhat arbitrary decisions that can potentially further alter our best fit. These decisions include the number and width of bins in the histograms and the number of days of shower activity analyzed. We conducted a crude sensitivity analysis to determine whether these sources of error are significant. We found that the total width of our histogram made a negligible difference in the results, so we keep that fixed at 10$^\circ$. However, the bin width and activity window produced variations in $\mu_1$ that were similar to those produced by the bootstrap alone.

We decided to incorporate all significant sources of error by conducting a combined random selection and parametric bootstrap in which each fit iteration:
\begin{enumerate}
    \item uses a shower duration that is randomly selected from the set of duration values that encompass at least 50 shower members but $\lesssim 90$\% of the estimated total number of shower members,
    \item uses a histogram bin size selected randomly from the interval [0.1$^\circ$, 0.3$^\circ$], and
    \item is performed on a random sample of the shower members.
\end{enumerate}
We again used the middle 68\% of bootstrap fit results to define our parameter uncertainties; this is equivalent to providing 1-$\sigma$ uncertainties for a normal distribution.

\subsubsection{The effects of radiant measurement uncertainties}
\label{sec:spread}

In general, we use a bootstrap approach to estimate the contributions that various parameters make to our fit uncertainties (Section~\ref{sec:bootstrap}). However, the formal uncertainties on the radiant measurements will systematically increase the measured radiant dispersion, and it is not possible to reverse this process using a parametric bootstrap.

Therefore, we instead measure the degree of perceived broadening that the radiant measurement uncertainties add to an assumed Rayleigh distribution. We determine this by simulating an idealized set of radiants that are evenly distributed about the nominal shower radiant (that is, $\Delta \lambda_\mathrm{ref} (\lambda_0)$, $\beta_\mathrm{ref}(\lambda_0)$). We then vary these simulated radiants by the measured radiant uncertainties -- we generate one simulated radiant per measured radiant -- and measure the difference between the two distributions. The left panel of Fig.~\ref{fig:spread} illustrates this process for an assumed Rayleigh distribution with a mode of 0.51$^\circ$. We see that, after applying our parametric bootstrap, the apparent mode is slightly larger.

We repeated this process for ``true'' Rayleigh distribution mode values ranging from 0.05$^\circ$ to 1$^\circ$. We fit a double Rayleigh distribution to the apparent distributions, although for values of $\mu_1$ greater than $\sim 0.1$, the second component of the best-fit double Rayleigh was very weak or nonexistent (that is, $\gamma \lesssim 0.1$). The relationship between the true mode and the mode of the main component of the best-fit double Rayleigh is shown in the right panel of Fig.~\ref{fig:spread}. We find that [1] the radiant uncertainties, if accurate, contribute very little to the measured spread and [2] we can obtain the true value of $\mu_1$ from the apparent value using a simple linear expression:
\begin{align}
    \mu_1 &= 1.029 \mu_{1, \, \textrm{app}} - 0.036^\circ ~ ~ (\text{for} ~ \mu_1 < 1)
    \label{eq:muapp1}
\end{align}
where $\mu_{1, \, \textrm{app}}$ is the apparent or measured value.

For larger values of $\mu_1$ ($\mu_1 \gtrsim 1$), the broadening of the peak due to radiant measurement uncertainty is undetectable. We fare the two regimes together as follows:
\begin{align}
    \mu_1 &= \begin{cases} 
      ~1.029 \mu_{1, \, \textrm{app}} - 0.036^\circ & 
        \mu_{1, \, \textrm{app}} < 1.2413^\circ \\
      ~\mu_{1, \, \textrm{app}} & 
        \mu_{1, \, \textrm{app}} \ge 1.2413^\circ
   \end{cases} \label{eq:muapp}
\end{align}
where the boundary value of $\mu_{1, \, \textrm{app}} = 1.2413^\circ = 0.036^\circ/(1.029-1)$ is simply the value for which the two parts of equation~\ref{eq:muapp} are equal. 
All values presented in Section~\ref{sec:results} incorporate this correction.

If the radiant uncertainties are underestimated, this effect could be larger. For instance, \cite{vida2021global} suggests that the uncertainties might be underestimated by as much as a factor of 2; if we multiply all uncertainties by 2, the slope of equation~\ref{eq:muapp1} increases to 1.067. However, this would still be a fairly minor source of error in our fit parameters.

\subsubsection{The correlation between radiant dispersion and shower duration}
\label{sec:dur}

Meteor showers contain particles of varying sizes and of varying ages. Small particles will typically be ejected from their parent comets at greater speeds and will be more affected by radiation pressure and Poynting-Robertson drag. Over time, the orbits of particles of all sizes will tend to diverge. If a meteoroid stream contains a component that is more evolved, we would expect this to be reflected in both the radiant dispersion and the activity duration. For instance, \cite{1998MNRAS.301..941J} found that the radiant dispersion of Perseid meteors was smaller during an outburst than it is in a typical year. A similar effect has been reported for the mass distribution within a shower; \cite{2000pimo.conf..112A} report a larger proportion of bright meteors near the peak of Taurid activity, particularly during outburst years.

\begin{figure*}
    \centering
    \includegraphics{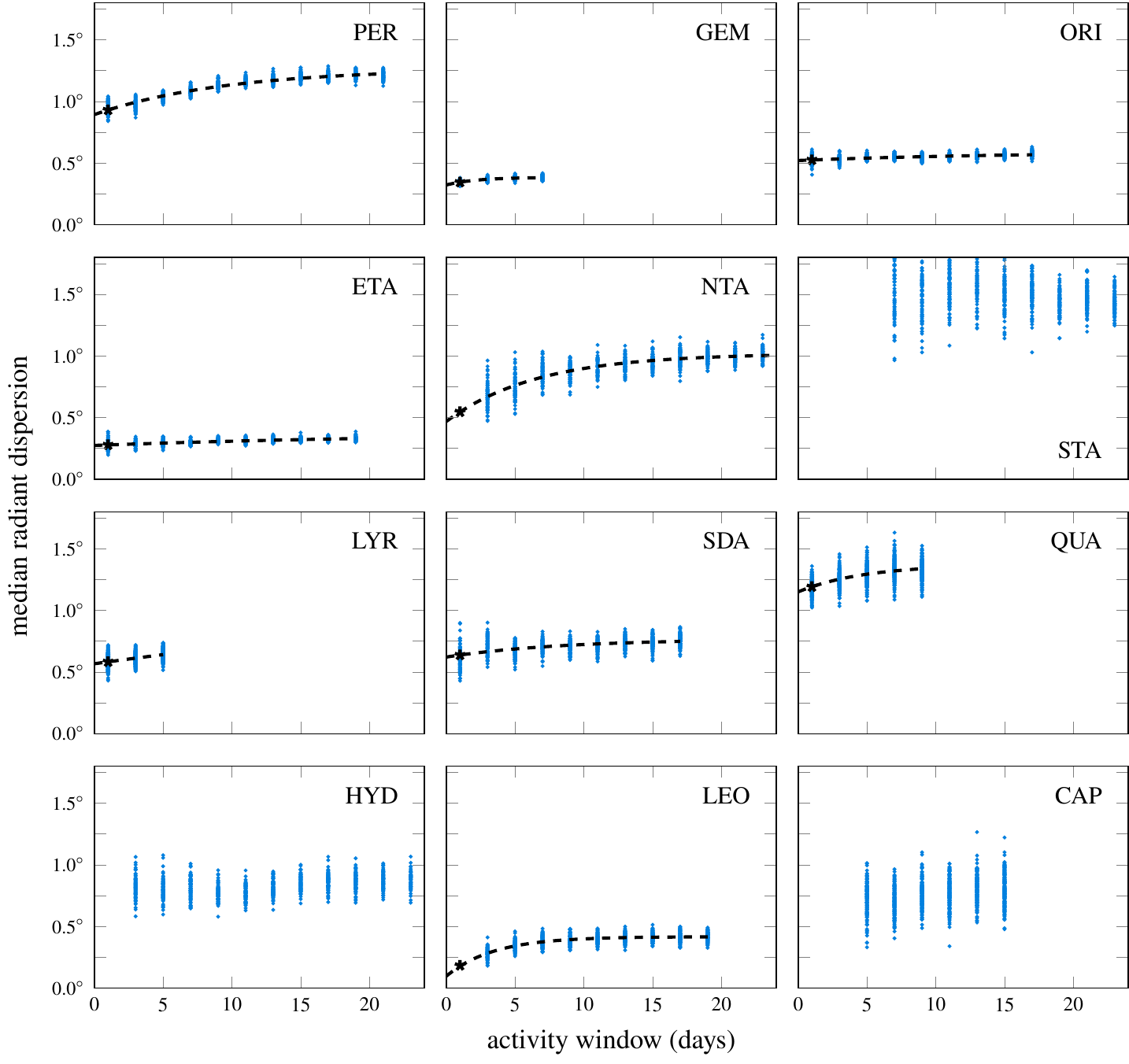}
    \caption{Bootstrap fit results (blue dots) for the median angular dispersion as a function of the width of the activity interval used for the fit. An exponential fit to the data is shown as a dashed line, and the extrapolated value corresponding to one night's worth of data is marked with a black star.}
    \label{fig:dur}
\end{figure*}

The number statistics for most of our showers do not permit us to control for this effect by restricting our analysis to one night of observations. However, our bootstrap uncertainty analysis does include several variations per shower of the number of days of data analyzed. We can use these bootstrap results to test whether there is a correlation between the duration of the analysis window and the derived radiant dispersion. Figure~\ref{fig:dur} shows such an analysis for all twelve meteor showers. 

Nine of our showers show a correlation between the duration of the activity window and the median dispersion. We see that, for these showers, the dispersion is lowest when the analysis window is shortest, and plateaus to a larger value for longer analysis windows. A simple linear fit does not capture this behavior; instead, we fit the following equation to the results:
\begin{align}
    \text{median} &= A + B \cdot C^{\text{dur}_\text{max}/2}
    \label{eq:abc}
\end{align}
where dur$_\text{max}$ is the width of the analysis window, $A$ and $B$ are positive constants, and $C$ is a constant between 0 and 1. This formula is a reasonable description of the relationship between activity window duration and median dispersion for the Perseids (PER), Geminids (GEM), Orionids (ORI), eta Aquariids (ETA), Northern Taurids (NTA), April Lyrids (LYR), Southern delta Aquariids (SDA), Quadrantids (QUA), and Leonids (LEO): data and trends are shown in Fig.~\ref{fig:dur}. The Southern Taurids (STA), sigma Hydrids (HYD), and alpha Capricornids (CAP) did not show clear signs of such a trend, but we also note that these three showers have the largest spread in bootstrap solutions.

\section{Results}
\label{sec:results}

This section summarizes our radiant dispersion measurements for the twelve showers we considered. Many of these results were depicted graphically in Section~\ref{sec:methods}, but here we provide additional details, compare our results with previous measurements, and discuss the implications for stream shape.

\subsection{Sun-centered ecliptic radiant drift}
\label{sec:driftres}

Most showers displayed a statistically significant drift in Sun-centered ecliptic radiant when we set our significance level at 0.05. Table~\ref{tab:drift} gives our best fit and corresponding uncertainties for the constants listed in equations~\ref{eq:driftl}--\ref{eq:driftv}. In those cases where the null hypothesis (i.e., that there is no correlation between the parameter and solar longitude) could not be ruled out at the 95\% confidence level, the best-fit slope appears in gray. We have still chosen to present these fit values, however, because they are a key component of our bootstrap uncertainty analysis.

{
\setlength{\tabcolsep}{10pt}
\begin{table*}
\centering
\begin{tabular}{crr@{\hskip 6pt}rr@{\hskip 6pt}rr@{\hskip 6pt}r}
    \hline
    \hline
    & 
        & \mcd{$\lambda_g - \lambda_\odot$} 
        & \mcd{$\beta_g$} 
        & \mcd{$v_g$} \\
    shower & \mcs{$\lambda_0$} 
        & \mcs{$b_\lambda$} & \mcs{$m_\lambda$} 
        & \mcs{$b_\beta$} & \mcs{$m_\beta$}
        & \mcs{$b_v$} & \mcs{$m_v$} \\
    code & \mcs{($^\circ$)} 
        & \mcs{($^\circ$)} &
        & \mcs{($^\circ$)} &
        & \mcs{(km s$^{-1}$)} & \mcs{(km s$^{-1}$ deg$^{-1}$)} \\ \hline
    PER & 140.10 
        & 283.531(41) &  0.068(14) 
        &  38.362(31) & -0.059(10) 
        &  58.680(36) & \insig{-0.011(11)} \\
    GEM & 262.20 
        & 207.995(12) & -0.092(17) 
        &  10.466(12) & -0.103(15) 
        &  33.881(22) &  0.088(29) \\
    ORI & 209.20
        & 246.670(25) & -0.242(09) 
        &  -7.557(17) &  0.063(06) 
        &  65.548(48) & -0.087(16) \\
    ETA &  45.12 
        & 293.813(19) & -0.254(06) 
        &   7.696(21) &  0.063(07) 
        &  65.508(48) &  0.058(13) \\
    NTA & 235.80 
        & 189.780(66) & -0.260(10) 
        &   2.613(45) &  0.014(06) 
        &  27.064(51) & -0.138(08) \\
    STA & 218.00 
        & 191.821(98) & -0.213(11) 
        &  -4.911(47) & -0.021(06) 
        &  27.192(96) & -0.066(10) \\
    LYR &  32.50 
        & 240.927(86) &  0.64(16)\hphantom{0} 
        &  56.639(42) & -0.22(10)\hphantom{0}
        &  46.657(54) &  0.233(72) \\
    SDA & 126.70 
        & 208.629(48) & -0.237(29) 
        &  -7.488(35) & -0.123(18) 
        &  40.419(66) & -0.196(29) \\
    QUA & 283.00 
        & 275.67(24)\hphantom{0} &  \insig{0.22(59)}\hphantom{0}
        &  63.458(56) &  \insig{0.17(13)}\hphantom{0}
        &  40.851(68) & \insig{-0.01(17)}\hphantom{0} \\
    HYD & 254.80 
        & 231.099(83) & -0.098(18) 
        & -16.421(53) & \insig{-0.001(14)} 
        &  58.766(94) & -0.059(22) \\
    LEO & 235.50 
        & 272.525(38) & -0.295(21) 
        &  10.225(43) & -0.169(21) 
        &  69.839(85) &  0.056(27) \\
    CAP & 125.48 
        & 179.27(14)\hphantom{0} & -0.359(56) 
        &   9.743(64) &  0.104(19) 
        &  22.364(69) & -0.158(23)
\end{tabular}
\caption{Solar longitude of peak activity (as observed by GMN cameras; $\lambda_0$) and linear fits for drift in Sun-centered ecliptic radiant ($\lambda_g - \lambda_\odot$, $\beta_g$) and geocentric speed ($v_g$) around this time. The fit parameters correspond to equations~\ref{eq:driftl}--\ref{eq:driftv}, where $m$ is used for slope and $b$ is used for the ``intercept'' (value when $\lambda_\odot = \lambda_0$). Slopes that lie within two $\sigma$ of zero appear in gray.}
\label{tab:drift}
\end{table*}
}

Eleven out of twelve showers exhibit a statistically significant drift in either $\lambda_g - \lambda_\odot$ or $\beta_g$, with one exception: the Quadrantids (QUA). However, the Quadrantids are also very brief in duration, which makes the detection of a drift difficult and also less critical for measuring radiant offsets.

\subsection{Nominal radiant dispersions}
\label{sec:nom}

Table~\ref{tab:nomfits} lists our nominal radiant dispersion fit parameters for each meteor shower. We include the best-fit parameters for equation~\ref{eq:pdf1} or equation~\ref{eq:pdf2}, depending on whether a Rayleigh or double Rayleigh provides a better fit to the data, and the corresponding mode and median dispersion. The last two columns provide the median dispersion from \cite{1970BAICz..21..153K} and an estimate of the median dispersion derived from \cite{1954MNRAS.114..229W}.

\begin{table*}
    \centering
    \begin{tabular}{cc@{\hskip 20pt}cccc@{\hskip 20pt}cccc}
        \hline \hline
        shower & $N$ & $\mu_1$ & $\mu_2$ & $\gamma$ & $c$
               & mode & median & \citeauthor{1970BAICz..21..153K} & \citeauthor{1954MNRAS.114..229W} \\
       & & ($^\circ$) & ($^\circ$) & & & ($^\circ$) & ($^\circ$) & median ($^\circ$) & adjusted ($^\circ$) \\
        \hline
        PER & 1445 & 0.77(10) & 1.53(31) & 0.39(14) & 0.009(12) & 0.84(08) & 1.14(11) & \hphantom{0}1.26 ($N = 254$) & \hphantom{0}1.02 ($N = 109$) \\
        GEM & 1279 & 0.26(03) & 0.53(06) & 0.35(12) & 0.001(01) & 0.28(02) & 0.38(02) & 0.49 ($N = 81$) & 0.33 ($N = 46$) \\
        ORI & 777 & 0.38(03) & 0.88(16) & 0.32(11) & 0.018(09) & 0.40(03) & 0.55(02) & 0.84 ($N = 30$) & 0.57 ($N = 28$) \\
        ETA & 441 & 0.21(03) & 0.49(07) & 0.37(11) & 0.018(05) & 0.22(02) & 0.32(02) & -- & --  \\
        NTA & 410 & 0.61(12) & 1.32(20) & 0.45(19) & 0.120(24) & 0.67(12) & 0.97(11) & 1.94 ($N = 25$) & 1.58 ($N = 25$) \\
        STA & 345 & 0.64(21) & 1.44(18) & 0.78(17) & 0.064(45) & 0.89(23) & 1.41(12) & 2.20 ($N = 46$) & 0.86 ($N = 47$) \\
        LYR & 333 & 0.41(06) & 1.09(18) & 0.38(14) & 0.020(08) & 0.43(06) & 0.64(05) & 0.25 ($N = 7$)\hphantom{0} & -- \\
        SDA & 285 & 0.52(16) & 1.28(48) & 0.20(30) & 0.043(12) & 0.54(12) & 0.70(06) & 1.41 ($N=27$) & 1.20 ($N=40$) \\
        QUA & 253 & 1.05(08) & -- & -- & 0.016(35) & 1.05(08) & 1.23(09) & 1.13 ($N = 16$) & -- \\
        HYD & 235 & 0.74(07) & -- & -- & 0.10(11)\hphantom{0} & 0.74(07) & 0.87(08) & -- & -- \\
        LEO & 170 & 0.16(06) & 0.49(09) & 0.65(21) & 0.036(27) & 0.18(07) & 0.38(05) & 0.32 ($N=9$) & 0.33 ($N=32$) \\
        CAP & 115 & 0.59(10) & -- & -- & 0.187(61) & 0.59(10) & 0.69(12) & -- & -- 
    \end{tabular}
    \caption{Radiant distribution fit parameters and their  corresponding median and mode for nine showers observed by GMN. Bootstrap uncertainties are provided in parentheses, and showers are sorted by decreasing number of observed meteors. The Quadrantids (QUA), sigma Hydrids (HYD), and alpha Capricornids (CAP) are best described by a single Rayleigh distribution and so have no values quoted for $\mu_2$ and $\gamma$. \protect\cite{1970BAICz..21..153K} median angular offset values are provided in the second-to-last column, and \protect\cite{1954MNRAS.114..229W} values, corrected using the factor in equation~\ref{eq:wwcorr} in order to obtain a value equivalent to the median dispersion, are provided in the last column.}
    \label{tab:nomfits}
\end{table*}

We found that the Rayleigh and double Rayleigh fits for most showers had very similar values for the median offset angle, so this is the most stable measure of the dispersion. However, \cite{2020MNRAS.494.2982M} found that the mode was a more appropriate choice for computing the effects of gravitational focusing, and therefore we provide both statistics.

We have included prior dispersion measurements from \cite{1970BAICz..21..153K} for several showers. However, this earlier work was based on many fewer meteors in most cases; for instance, we have almost 50 times as many Lyrid radiants on which to base our analysis. Of the showers considered by this earlier work, only the Perseids would have satisfied our threshold for analysis ($N > 100$). We note that in this case, our measurement agrees with that of \cite{1970BAICz..21..153K} in that they differ by about 1~$\sigma$.

We have also included radiant dispersion measurements from \cite{1954MNRAS.114..229W}, adjusted to account for differences in methodology. \citeauthor{1954MNRAS.114..229W} cite mean rather than median deviations; we multiply by a factor of $\sqrt{\pi/4 \ln 2}$ to account for the difference between the median and the mean of a Rayleigh distribution. We multiply by an additional factor of $\sqrt{2}$ to account for the fact that \cite{1954MNRAS.114..229W} measures only one of two radiant offset dimensions. Thus, the total correction factor is:
\begin{align}
    \sqrt{\frac{\pi}{2 \ln 2}} &\simeq 1.5
    \label{eq:wwcorr}
\end{align}
This factor has been applied to obtain the values in the last column of Table~\ref{tab:nomfits}. After applying this correction, the \cite{1954MNRAS.114..229W} results for the Perseids, Geminids, Orionids, and Leonids are within one-to-two $\sigma$ of our measurements; however, the Taurid and Southern delta Aquariid values are substantially different. Discrepancies between our values, those of \cite{1970BAICz..21..153K}, and those of \cite{1954MNRAS.114..229W} indicate that the Taurids in particular may be difficult to characterize. Additionally, all three studies measure disparate dispersions for the Northern and Southern Taurids, but do not all measure a larger dispersion for the same branch. Both \cite{1970BAICz..21..153K} and this work measure larger dispersions for the Southern Taurids, while \cite{1954MNRAS.114..229W} measures a larger dispersion for the Northern Taurids.

Finally, we note that in \cite{2020MNRAS.494.2982M}, we obtained a crude measurement of the radiant dispersion of Orionid and Perseid meteors using the first 18 months out of the 30 months of data used in this work. This earlier attempt does not subtract sporadic contamination or account for radiant measurement uncertainties, and fits a single Rayleigh distribution to a cropped portion of the radiant dispersion data using a maximum likelihood estimation algorithm. The inclusion of a larger data set and refinement of our fitting algorithms resulted in a measured mode that is smaller for both showers: we measure the mode of the Perseid offsets as $0.84 \pm 0.08^\circ$ instead of 1.02$^\circ$, and for the Orionids we now obtain $0.40 \pm 0.03^\circ$ instead of 0.52$^\circ$

\subsection{Radiant dispersions on the most active night}
\label{sec:1night}

The bootstrap fits for nine of our showers exhibited a correlation between the number of days of shower activity analyzed and the median radiant dispersion (see Section~\ref{sec:dur} and Fig.~\ref{fig:dur}). We use these fits to estimate the median radiant dispersion on the night that yielded the largest number of GMN observations; these values are listed in Table~\ref{tab:1night}.

We note that equation~\ref{eq:abc} was concocted to resemble the behavior seen in Fig.~\ref{fig:dur} and has no physical motivation. Therefore, the numbers in Table~\ref{tab:1night} should be considered crude estimates of how tight the dispersion may be on the most active night, and are presented without formal uncertainties.

\subsection{Shower duration and radiant dispersion}

One might expect that both radiant dispersion and shower duration probe the spread of orbital elements within a meteor shower and thus could be correlated. \cite{1954MNRAS.114..229W} reported such a correlation for measurements of the Draconids, Leonids, Geminids, Orionids, Perseids, delta Aquariids, and Northern and Southern Taurids.
We test for this correlation in our data by comparing our median radiant dispersions with the duration of each shower.

We do not measure shower duration from this data, as we have not corrected for issues such as gaps in data collection due to cloudy weather. Instead, we will rely on existing measures of the showers' activity profiles, which are often modeled as a double exponential function: \citep{1994A&A...287..990J,Moorhead2017}:
\begin{align}
    \text{ZHR} &= \text{ZHR}_0 \times 
    \begin{cases} 
        10^{+B_p (\lambda_\odot - \lambda_0)} & \lambda_\odot < \lambda_0 \\
        10^{-B_m (\lambda_\odot - \lambda_0)} & \lambda_\odot \ge \lambda_0
    \end{cases}
\end{align}
Here, ZHR refers to the zenithal hourly rate, ZHR$_0$ is the rate at the shower's peak, $\lambda_0$ is the solar longitude corresponding to the peak, and $B_p$ and $B_m$ are parameters that govern how steeply activity rises in advance of the peak or drops after it.

\begin{table}
    \centering
    \begin{tabular}{cccc}
        \hline \hline
        shower & nom.~window & nom.~median & median$_1$ \\
        code & (days) & ($^\circ$) & ($^\circ$) \\
        \hline
        PER & 9 & 1.14(11) & 0.93 \\
        GEM & 3 & 0.38(02) & 0.34 \\
        ORI & 7 & 0.55(02) & 0.53 \\
        ETA & 9 & 0.32(02) & 0.28 \\
        NTA & 15 & 0.97(11) & 0.55 \\
        STA & 19 & 1.41(12) & -- \\
        LYR & 3 & 0.64(05) & 0.58 \\
        SDA & 5 & 0.70(06) & 0.64 \\
        QUA & 1 & 1.23(09) & 1.19 \\
        HYD & 15 & 0.87(08) & -- \\
        LEO & 7 & 0.38(05) & 0.18 \\
        CAP & 7 & 0.69(12) & -- 
    \end{tabular}
    \caption{Estimated median meteor shower radiant dispersion on the night when the largest number of meteors were detected by GMN (median$_1$). We compare these values with the median angular offset of our nominal fit (taken from Table~\ref{tab:nomfits}) and provide the duration of the activity window used in our nominal fit.}
    \label{tab:1night}
\end{table}

The full duration at half maximum (FDHM) of such an activity curve has the following analytic solution:
\begin{align}
    \text{FDHM} &= \left( 
        \frac{1}{B_p} + \frac{1}{B_m}
    \right) \cdot \log_{10} 2 \label{eq:fwhm}
\end{align}
Some showers, such as the Geminids and Perseids, are better described by a superposition of two double-exponential curves; in these cases, we solve for the FDHM numerically. In general, we take the $B_p$ and $B_m$ parameters from \cite{Moorhead2017}, except in the case of the Geminids.  We conducted spot-checks of the radar profiles against publicly available optical meteor fluxes\footnote{https://www.imo.net/members/imo\_live\_shower}\footnote{https://meteorflux.org} and find that, except for the Geminids, the \cite{Moorhead2017} profiles are also a good fit to optical data. Because optical Geminid activity tends to be briefer than radar Geminid activity, we use the activity parameters computed by \cite{1994A&A...287..990J} using visual Geminid observations.

\begin{figure}
    \centering
    \includegraphics{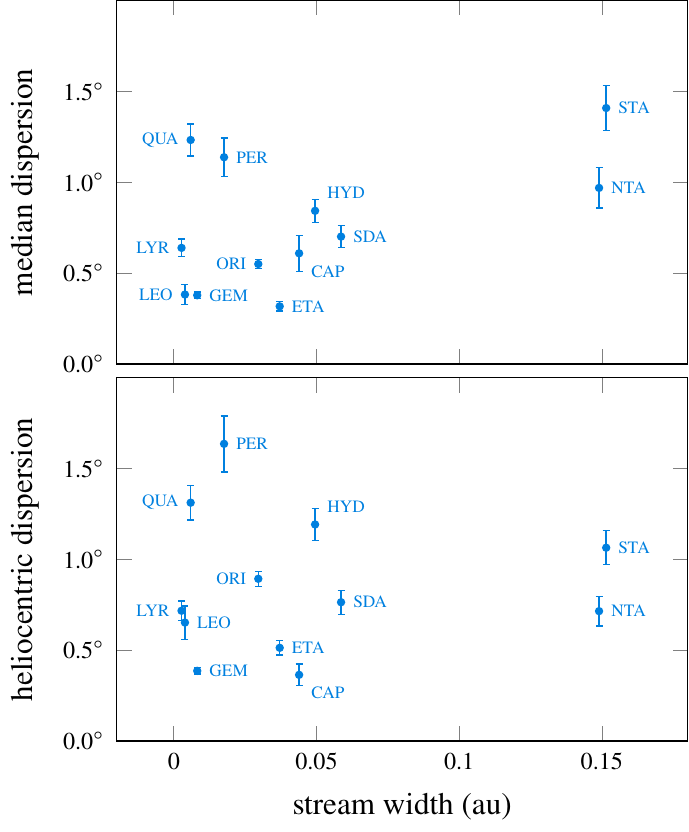}
    \caption{A comparison of the median geocentric radiant dispersion (top) and heliocentric radiant dispersion (bottom) with a measure of the stream's width derived from its duration.}
    \label{fig:fwhm}
\end{figure}

We use equation~24 of \cite{1970BAICz..21..153K} to compute the ``width'' of a meteoroid stream from its duration:
\begin{align}
    S &= r_h \tan \left( \text{FDHM} / 2 \right) \sin \epsilon_h
\end{align}
where $r_h \simeq 1$~au is the heliocentric distance at the time of observation and $\epsilon_h$ is the angle between the apex direction and the heliocentric radiant. In Fig.~\ref{fig:fwhm}, we compare this quantity with both the median geocentric radiant dispersion and the equivalent heliocentric radiant dispersion, which is given by
\begin{align}
    \Theta_{\text{med}, \, h} &= \tan^{-1} \left( 
        \frac{v_g}{v_h} \tan \Theta_{\text{med}, \, g}
    \right) 
\end{align}
where $\Theta_\text{med}$ is used to indicate the median angular offset and $v_h$ is the heliocentric speed of the stream. Note that we plot the median dispersion within our nominal window for each shower (see Table~\ref{tab:1night} for these durations). The radiant dispersion for the entire stream may be a little broader than that for our nominal windows, as indicated by Fig.~\ref{fig:dur}, but the difference is small for most showers.

We do not observe a correlation between either geocentric or heliocentric radiant dispersion and stream width/duration. A Spearman rank-order correlation test fails in both cases. Thus, we do not reproduce the result of \cite{1954MNRAS.114..229W}. This is likely due to the difference in methodology in determining the shower duration. \cite{1954MNRAS.114..229W} note that their durations are highly uncertain and are based on the elapsed time between the first observed meteor attributed to a shower and the last; this quantity will depend on the level of activity as well as the shape of the time profile. For instance, their Perseid duration is approximately 27 days, while ours is about 2 days. Furthermore, they lack dispersion measurements for the Quadrantids and eta Aquariids, whose placements in Fig.~\ref{fig:fwhm} tend to inhibit a correlation.

This has several implications: first, shower duration cannot be used to predict the radiant dispersion of a shower. Second, variations in the ratio of shower duration to radiant dispersion may indicate that the cross section of some streams are elliptical or asymmetrical \citep{2005A&A...439..761V}. These two quantities could also potentially be used as an independent constraint on dynamical models of meteoroid streams.

\section{Conclusions}

We have presented measurements of Sun-centered ecliptic radiant drift and radiant dispersions of meteors belonging to twelve major showers: the Perseids, Geminids, Orionids, eta Aquariids, Northern Taurids, Southern Taurids, April Lyrids, Southern delta Aquariids, Quadrantids, sigma Hydrids, Leonids, and alpha Capricornids. These are the first radiant dispersion measurements for the eta Aquariids, sigma Hydrids, and alpha Capricornids. Our measurements are based on high quality meteor trajectories from the Global Meteor Network which, critically, include formal measurement uncertainties.

Meteor radiants are known to drift in right ascension and declination over time, and linear approximations of these drifts are published in observer tables. Meteor astronomers therefore often use Sun-centered ecliptic longitude and latitude to describe the position of a meteor shower, as shower radiants tend to be more stationary in this non-inertial coordinate system. However, we have measured small but statistically significant drifts in Sun-centered ecliptic radiant for eleven of the twelve showers considered here. Ten showers also show signs of a drift in geocentric velocity. 

While the distribution of radiant offsets is similar to a Rayleigh distribution \citep[as established by][]{2020MNRAS.494.2982M}, most showers show a slight-to-moderate excess of meteors towards the tail-end of the distribution. We have found that a double Rayleigh distribution is better able to replicate this shape. We have not offered a physical justification for the double Rayleigh distribution, but possible explanations are that streams consist of meteoroid populations of different sizes or ages that have varying dispersions, or that the stream cross-section is not axisymmetric.

We have placed particular emphasis on determining the smallest plausible dispersion; our motivation in doing so is to enable conservativism in applying these results to spacecraft risk predictions \citep{2020MNRAS.494.2982M}. While radiant measurement errors have a modest effect on the radiant dispersion (see Section~\ref{sec:spread}), the duration of the analysis window appears to have a significant effect. We have estimated the one-night radiant dispersion for our showers and find that the Northern Taurids and Leonids could have a significantly more compact radiant near the time of peak activity.

Our results suggest that radiant dispersion measurements could offer new constraints on meteoroid stream structure. Unlike \cite{1954MNRAS.114..229W}, we do not observe a simple linear correlation between radiant dispersion and stream width (as determined from shower duration). This opens up the possibility that some of these meteoroid streams may be elliptical in cross-section. This is supported by other studies: for instance, both observations \citep{vida2020phdthesis} and models \citep{egal2020modeling} of the Orionid meteor shower indicate a complex radiant structure. An apparently asymmetric Orionid radiant distribution is also presented in Fig.~1 of \cite{2020MNRAS.494.2982M}. Thus, we encourage additional characterizations of meteor shower radiant distributions. Future improvements could include investigations of whether this dispersion varies with limiting meteor magnitude, and the degree of asymmetry present in the stream cross section (a somewhat tricky measurement for a group of points that slowly drifts across a sphere). We also suggest that measured radiant distributions be compared with those predicted by meteoroid stream models.

\section*{Acknowledgments}

This work was supported in part by NASA Cooperative Agreement 80NSSC18M0046, by the Natural Sciences and Engineering Research Council of Canada, and by Jacobs contract 80MSFC18C0011. Initial funding for the 16~mm GMN stations was provided by the Istria County Association of Technical Culture.

The authors would like to thank the following GMN station operators and contributors whose stations provided the data used in this work (in alphabetical order): Victor Acciari, Alexandre Alves, {\v Z}eljko Andrei{\' c}, Georges Attard, Roger Banks, Hamish Barker, Ricky Bassom, Richard Bassom, Jean-Philippe Barrilliot, Dr. Ehud Behar, Josip Belas, Alex Bell, Serge Bergeron, Arie Blumenzweig, Ventsislav Bodakov, Ludger B\"{o}rgerding, Claude Boivin, Bruno Bonicontro, Fabricio Borges, Dorian Bo\v{z}i\v{c}evi\'{c}, Martin Breukers, John Briggs, Laurent Brunetto, Tim Burgess, Peter Campbell-Burns, Pablo Canedo, Seppe Canonaco, Jose Carballada, Gilton Cavallini, Brendan Cooney, Edward Cooper, Dino {\v C}aljku{\v s}i{\' c}, Tim Claydon, Manel Colldecarrera, Christopher Curtis, Ivica {\' C}ikovi{\' c}, J{\" u}rgen D{\"o}rr, Chris Dakin, Alfredo Dal'Ava J{\'u}nior, Steve Dearden, Christophe Demeautis, Paul Dickinson, Ivo Dijan, Tammo Jan Dijkema, Pieter Dijkema, Stacey Downton, Zoran Dragi\'{c}, Jean-Paul Dumoulin, Garry Dymond, Robin Earl, Ollie Eisman, Carl Elkins, Peter Eschman, Rick Fischer, Richard Fleet, Jim Fordice, Mark Gatehouse, Megan Gialluca, Jason Gill, Hugo Gonz\'{a}lez, Philip Gladstone, Uwe Gl\"{a}ssner, Nikola Gotovac, Neil Graham, Bob Greschke, Sam Green, Daniel J. Grinkevich, Larry Groom, Tioga Gulon, Dominique Guiot, Margareta Gumilar, Pete Gural, Kees Habraken, Erwin Harkink, Ed Harman, Tim Havens, Richard Hayler, Alex Hodge, Bob Hufnagel, Russell Jackson, Jean-Marie Jacquart, Ron James Jr., Ilya Jankowsky, Klaas Jobse, Dave Jones, Vladimir Jovanovi{\'c}, Milan Kalina, Jonathon Kambulow, Richard Kacerek, Steve Kaufman, Alex Kichev, Jean-Baptiste Kikwaya, Zoran Knez, Dan Klinglesmith, Danko Ko{\v c}i{\v s}, Korado Korlevi{\'c}, Stanislav Korotkiy, Josip Krpan, Zbigniew Krzeminski, Patrik Kuki\'{c}, Reinhard K\"{u}hn, Ga\'{e}tan Laflamme, David Leurquin, Anton Macan, John Maclean, Igor Macuka, Mirjana Malari{\'c}, Nedeljko Mandi{\'c}, Bob Marshall, Jos\'{e} Luis Martin, Colin Marshall, Andrei Marukhno, Keith Maslin, Bob Massey, Damir Matkovi\'{c}, Michael J. Mazur, Sergio Mazzi, Alex McConahay, Robert McCoy, Charlie McCromack, Mark J.M. McIntyre, Aleksandar Merlak, Filip Mezak, Pierre-Michael Micaletti, Matej Mihel{\v c}i{\'c}, Simon Minnican, Wullie Mitchell, Nick Moskovitz, Gene Mroz, Brian Murphy, Carl Mustoe, Przemek Naga{\'n}ski, Jean-Louis Naudin, Damjan Nemarnik, Colin Nichols, Zoran Novak, Michael O'Connell, Washington Oliveira, Thiago Paes, Carl Panter, Lovro Pavleti\'{c}, Filip Parag, Igor Pavleti\'{c}, Richard Payne, Pierre-Yves Pechart, Enrico Pettarin, Alan Pevec, Patrick Poitevin, Pierre de Ponthière, Alex Pratt, Miguel Preciado, Chuck Pullen, Dr. Lev Pustil'nik, Chris Ramsay, Danijel Reponj, David Robinson, Martin Robinson, Heriton Rocha, Herve Roche, Adriana Roggemans, Paul Roggemans, Alex Roig, James Rowe, Dmitrii Rychkov, Michel Saint-Laurent, Jason Sanders, Rob Saunders, William Schauff, Ansgar Schmidt, Jay Shaffer, Jim Seargeant, Ivica Skoki\'{c}, Dave Smith, Tracey Snelus, James Stanley, Peter Stewart, William Stewart, Bela Szomi Kralj, Ian Pass, Rajko Su\v{s}anj, Damir \v{S}egon, Marko \v{S}egon, Jeremy Taylor, Yakov Tchenak, Eric Toops, Steve Trone, Wenceslao Trujillo, Paraksh Vankawala, Martin Walker, Bill Wallace, Didier Walliang, Jacques Walliang, Christian Wanlin, Tom Warner, Urs Wirthmueller, Steve Welch, Alexander Wiedekind-Klein, John Wildridge, Bill Witte, Stephane Zanoni, and Dario Zubovi\'{c}.

\section*{Data availability}

The data used in this article are available in the Global Meteor Network database at https://globalmeteornetwork.org/data/. Readers wishing to replicate our results should select those trajectories with dates between 10 Dec 2018 and 22 May 2021 (inclusive).

\bibliographystyle{mnras}
\bibliography{main} 


\appendix
\section{Residual analysis}

\subsection{Radiant drift}
\label{sec:resid1}

\begin{figure*}
    \centering
    \includegraphics{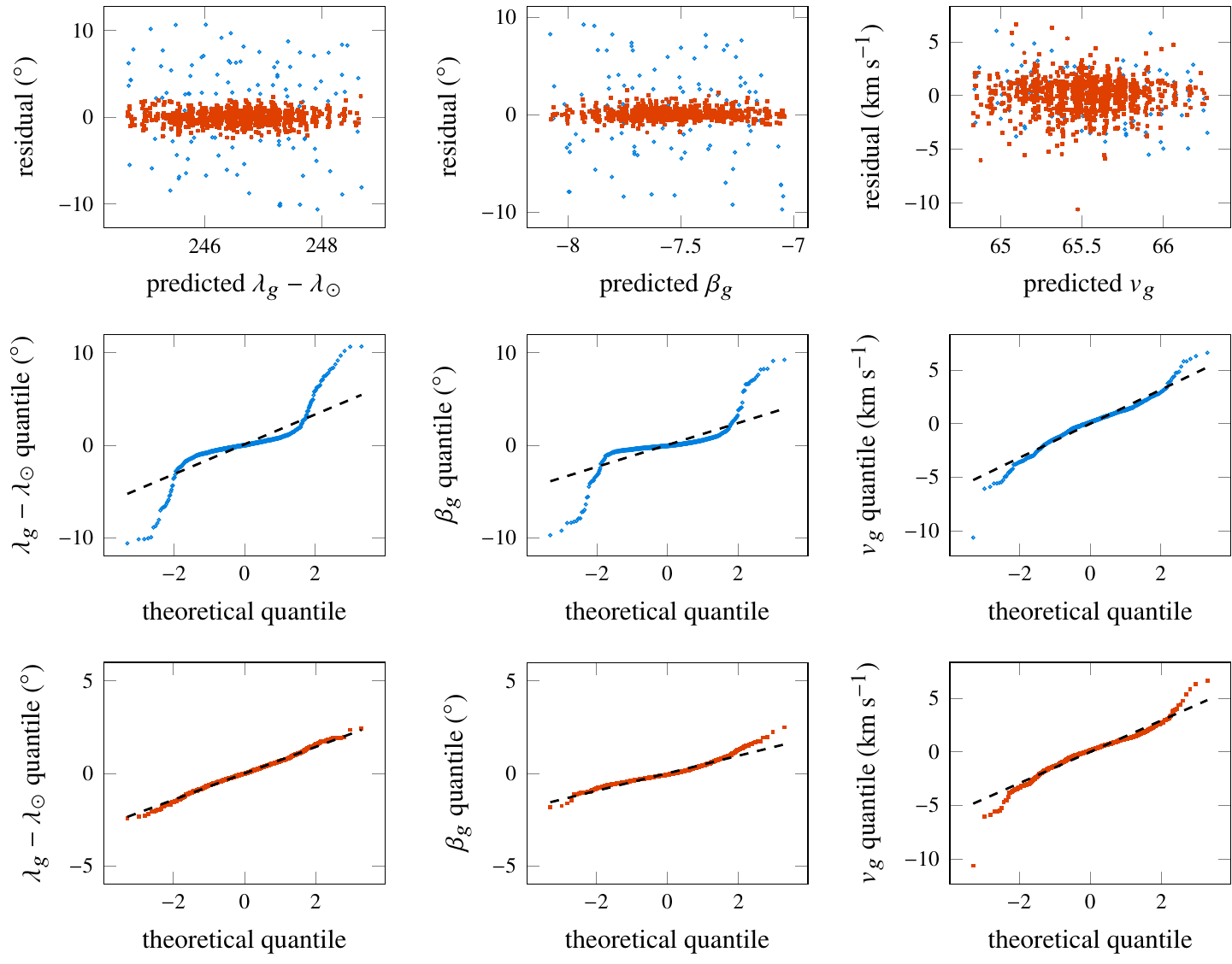}
    \caption{Residuals corresponding to our best-fit linear drifts in Orionid radiant and speed. The top row presents residuals-vs-fits plots for sun-centered ecliptic longitude (left), ecliptic latitude (center), and geocentric speed (right). Those meteors that lie more than 2.5$^\circ$ from the best-fit lines appear in blue. The middle row presents normal probability plots of the residuals, including all data; the bottom row presents normal probability plots of the residuals, including only points that lie within 2.5$^\circ$ of the best-fit lines.}
    \label{fig:linear_resids}
\end{figure*}

In Section~\ref{sec:drift}, we fit linear trends to the value of a shower's sun-centered ecliptic longitude, latitude, and geocentric speed as a function of solar longitude. Figure~\ref{fig:linear_fits} presents the results for the Orionid shower; Fig.~\ref{fig:linear_resids} presents the residuals corresponding to this fit.

We can see from the top row of Fig.~\ref{fig:linear_resids} that there are no obvious signs of non-linearity in the data, nor are there signs of unequal variance or residual correlations. Thus, three of the four conditions for using least-squares fitting appear to be satisfied. However, the normal probability plot in the middle row of Fig.~\ref{fig:linear_resids} indicates that the residuals are not normally distributed.

The distribution of the residuals of the radiant longitude and latitude appear to have heavier tails than a normal distribution would. This is because there are two populations contributing to the distribution: the meteor shower and the sporadic background. We find that if we clip the data to those meteors that lie within 2.5$^\circ$ of our nominal best-fit radiant drift line (which appear in red in the scatter plots of Fig.~\ref{fig:linear_resids}), the resulting residual distribution much more closely resembles a normal distribution (see bottom row of Fig.~\ref{fig:linear_resids}). This cutoff appears to work well for all showers considered in this paper, and motivates our clip-and-iterate approach to fitting linear trends to the radiant and speed drifts.

\subsection{Radiant offset distribution}
\label{sec:resid2}

\begin{figure*}
    \centering
    \includegraphics[width=0.75\linewidth]{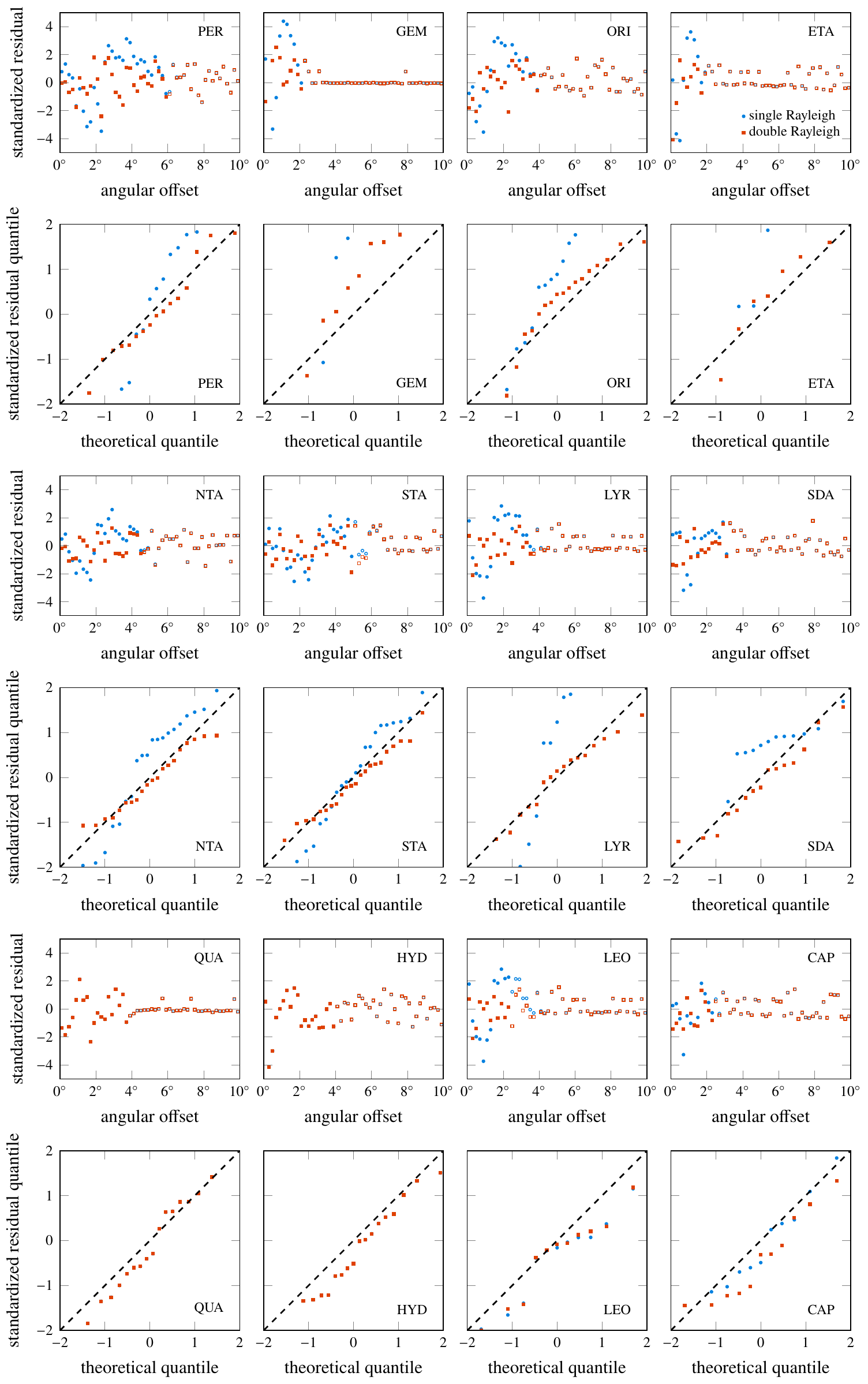}
    \caption{Standardized residuals (that is, residuals divided by one-$\sigma$ uncertainties) corresponding to our best-fit radiant offset dispersion for all analyzed showers. The top row of each set presents the standardized residuals as a function of the measured angular offset for both single and double Rayleigh fits. The bottom row presents normal probability plots of the standardized residuals; we mark unity with a black dashed line. Residuals included in the normal probability plots and reduced $\chi^2$ values appear as filled, rather than open, points.}
    \label{fig:pdf_resids}
\end{figure*}

In Section~\ref{sec:fit}, we fit Rayleigh and double Rayleigh distributions to noise-subtracted histograms of shower member radiant offsets. In Fig.~\ref{fig:pdf_resids}, we display residuals for each shower relative to both single and double Rayleigh fits. 

Figure~\ref{fig:pdf_resids} differs from Fig.~\ref{fig:linear_resids} in several ways: [1] we present standardized residuals (i.e., $(f_i - \hat{f}_i)/\sigma_{f_i}$) so that readers can easily see the deviation of the residuals from zero in units of $\sigma_{f_i}$; [2] we present residuals-vs-predictor plots rather than residuals-vs-fits so that readers can see how the standardized residuals have smaller variances when the angular offset is large; and [3] the dashed black lines in the normal probability plots are not fits to the data but instead correspond to unity, since we would ideally expect the standardized residuals to follow a standard normal distribution.

The unequal variance in standardized residuals between the core of the angular offset distribution and its tail is due to the limitations of how we estimate the uncertainty associated with empty bins (see equation~\ref{eq:wac}). The assignment of a constant uncertainty to empty bins out in the tail of the distribution results in over-large uncertainties and thus too-small standardized residuals. We cannot remove the tail from our fitting algorithm: if anything, we are under-weighting the tail in our fits. However, the standardized residuals in the tail will not be useful for assessing the goodness of fit or the normality of residuals; inclusion of the empty bins will drive down the reduced chi-squared statistic artificially. Thus, for each shower we locate the minimum value of $i$ for which $f_i \le 0$ (see equation~\ref{eq:fi}) and use only those values left of this boundary (which we will call $i_\mathrm{max}$) to compute the normal probability plot of the residuals and the reduced chi-squared statistic:
\begin{align}
    \chi^2_\mathrm{red} &= \frac{\sum_{i=1}^{i_\mathrm{max}-1} (f_i - \hat{f}_i)^2/\sigma^2_{f_i}}{i_\mathrm{max} - 1 - n_\mathrm{par}}
\end{align}
where $n_\mathrm{par} = 2$ for a Rayleigh distribution and 4 for a double Rayleigh distribution. Table~\ref{tab:chi2red} provides reduced chi-squared statistics for all fits and showers.

\begin{table}
    \centering
    \begin{tabular}{cccc}
    \hline \hline
    shower & $\chi^2_{\mathrm{red}, \, 1}$ & $\chi^2_{\mathrm{red}, \, 2}$ & evidence for \\
    code & & & double Rayleigh \\ \hline
    PER & 3.90 & 0.93 & strong \\
    GEM & 7.13 & 1.27 & strong \\
    ORI & 4.88 & 1.10 & strong \\
    ETA & 5.73 & 1.13 & strong \\
    NTA & 1.65 & 0.64 & weak \\
    STA & 1.73 & 1.00 & weak \\
    LYR & 3.61 & 1.35 & strong \\
    SDA & 2.26 & 1.01 & strong \\
    QUA & 1.06 & 1.22 & none \\
    HYD & 1.81 & 1.59 & none \\
    LEO & 3.66 & 0.83 & strong \\
    CAP & 2.95 & 2.22 & none
    \end{tabular}
    \caption{Reduced chi-squared statistic corresponding to both single and double Rayleigh fits ($\chi^2_{\mathrm{red}, \, 1}$ and $\chi^2_{\mathrm{red}, \, 2}$, respectively) to the angular offset distribution of each shower analyzed. The last column provides a qualitative assessment of the strength of the evidence in favor of a double Rayleigh fit over a single Rayleigh fit.}
    \label{tab:chi2red}
\end{table}

Using the rule of thumb that $\chi_\mathrm{red}^2 > 1$ indicates a poor fit and $\chi_\mathrm{red}^2 \sim 1$ indicates a reasonable fit, we find that a double Rayleigh distribution provides a much better fit for seven showers (PER, GEM, ORI, ETA, LYR, SDA, and LEO) and a somewhat better fit for the Taurids (NTA and STA); readers may come to the same conclusion simply by reviewing Fig.~\ref{fig:pdfs}. Three showers (QUA, HYD, and CAP) show no evidence of being better described by a double Rayleigh, although we note that these three are among the four least-populated showers in our data set.

Even with the exclusion of the tail of the distribution, some showers show signs of lack-of-fit or non-normality in the residuals. For instance, the Orionid residuals trend downward as the angular offset approaches zero, and the data points in the normal probability plot curve away from the unity line. In other cases, such as the Geminids, there are too few non-tail bins for us to test the normality of the residuals. Thus, it may be productive to revisit the shape of the radiant offset distribution of these showers after additional years of data are collected.

\bsp	
\label{lastpage}
\end{document}